\documentclass[twocolumn,showpacs,preprintnumbers,amsmath,amssymb,prl,superscriptaddress]{revtex4}
\usepackage{graphicx}
\usepackage{epstopdf}
\usepackage{dcolumn}
\usepackage{bm}
\usepackage{amsmath}
\usepackage{amssymb}
\usepackage{mathrsfs}
\usepackage{amsfonts}
\usepackage{times}
\usepackage{subfigure}
\usepackage{amstext}
\usepackage[colorlinks=true, citecolor=blue]{hyperref}
\usepackage[dvipsnames]{xcolor}

\newcommand{\beqa}{\begin{eqnarray}}
\newcommand{\eeqa}{\end{eqnarray}}

\newcommand{\bra}[1]{\langle #1|}
\newcommand{\ket}[1]{|#1\rangle}

\newcommand{\Cos}[1]{\cos\left( #1\right)}
\newcommand{\Sin}[1]{\sin\left( #1\right)}
\newcommand{\phidot}{\dot \phi}
\newcommand{\varphidot}{\dot \varphi}
\newcommand{\eff}{{\rm eff}}

\begin{document}
\title{Ultrastrong coupling regime of non-dipolar light-matter interactions}
\author{Simone Felicetti}
\affiliation{Laboratoire Mat\'eriaux et Ph\'enom\`enes Quantiques,
Universit\'e Paris Diderot-Paris 7 and CNRS, \\ B\^atiment Condorcet, 10 rue
Alice Domon et L\'eonie Duquet, 75205 Paris Cedex 13, France}
\author{Myung-Joong Hwang}
\affiliation{Insitut f\"ur Theoretische Physik and IQST, Albert-Einstein-Allee 11, Universit\"at Ulm, D-89069 Ulm, Germany}
\author{Alexandre Le Boit\'e}
\affiliation{Laboratoire Mat\'eriaux et Ph\'enom\`enes Quantiques,
Universit\'e Paris Diderot-Paris 7 and CNRS, \\ B\^atiment Condorcet, 10 rue
Alice Domon et L\'eonie Duquet, 75205 Paris Cedex 13, France}

\begin{abstract}
We present a circuit-QED scheme which allows to reach the ultrastrong coupling regime of a nondipolar interaction between a single qubit and a quantum resonator. We show that the system Hamiltonian is well approximated by a two-photon quantum Rabi model and propose a simple scattering experiment to probe its fundamental properties. In particular, we identify a driving scheme that reveals the change in selection rules characterizing the breakdown of the rotating-wave approximation and the transition from strong to ultrastrong two-photon interactions. Finally, we show that a frequency crowding in a narrow spectral region is observable in the output fluoresce spectrum as the coupling strength approaches the collapse point, paving the way to the direct observation of the onset of the spectral collapse in a solid-state device.
\end{abstract}

\maketitle
In cavity quantum electrodynamics, the interaction between photons and atoms is often described within the dipolar approximation \cite{Cohen-Tannoudji:1989}, leading to linear (single-photon) interaction terms, as exemplified by the celebrated Jaynes-Cummings and quantum Rabi models. Within this framework, considerable efforts have been made in the last decades to control and increase the light-matter coupling strength in various cavity QED experiments. The strong coupling regime where the coupling strength is larger than any dissipation rate has been demonstrated in atomic cavity QED \cite{Rempe:1987}, semiconductor nanostructures \cite{Reithmaier:2004, Peter:2005} and superconducting circuits \cite{Wallraff:2004}, leading to the observation of genuine quantum effects such as sub-Poissonian photon statistics \cite{Imamoglu:1997, Birnbaum:2005, Bozyigit:2010, Lang:2011, Hoffman:2011}. More recent progress has also made it possible to reach the so-called ultrastrong coupling regime where the coupling strength becomes comparable or even larger than the cavity frequency  \cite{Devoret:2007, Bourassa:2009, Todorov:2010, Niemczyk:2010, Forn-Diaz:2010, Nataf:2011, Forn-Diaz:2016, Forn-Diaz:2017,Yoshihara:2017}.

In this context, two-photon interaction processes (e.g. processes involving the simultaneous creation of one atomic excitation and absorption of two cavity photons) have so far been realized using second- or higher-order effects of the dipolar interaction in driven systems and therefore limited to extremely small coupling strengths~\cite{DiPiazza:2012, Hamsen:2017}.
However, a variety of novel physical phenomena emerges when the two-photon interaction (TPI) reaches the ultrastrong coupling regime. 
 In particular, a collapse of the discrete energy spectrum into a continuous band has been predicted in the ultrastrong coupling regime of various two-photon generalizations of the quantum Rabi model \cite{Travenec:2012, Albert:2011, Chen:2012, Duan:2016}. In the many-body limit, the TPI leads to a rich interplay between the spectral collapse and the superradiant phase transition~\cite{Garbe:2017, Chen:2018}.
These considerations prompted various efforts to design quantum simulators of TPI models in different atomic platforms \cite{Felicetti:2015, Puebla:2017,Schneeweiss:2017}. However, the implementation of a genuine TPI, where the coupling is not mediated by an external drive, requires an interaction more complex than dipolar. As recently shown, superconducting circuits are promising platforms for the design of such nondipolar interactions ~\cite{Felicetti:2018}.

In this letter, we show that fundamental quantum optical phenomena due to an ultrastrong nondipolar light-matter interaction can be observed with current circuit-QED technology. To this end, we propose and analyze a device that realizes the two-photon quantum Rabi model in the nonperturbative USC regime. We characterize the circuit response under coherent driving for increasing values of the coupling strength of the genuine TPI. The transition from strong to ultrastrong coupling is witnessed by the appearance of additional peaks in the fluorescence spectrum resulting from a change in selection rules due to the breakdown of the rotating wave approximation. In addition, higher-order photon correlations reveal the abrupt disappearance of nonlinear effects such as the two-photon blockade for specific coupling strengths in the USC regime. Finally we show that the output field bears a clear signature of the spectral collapse.

\paragraph{Circuit scheme}
\begin{figure}[t]
\centering
\includegraphics[angle=0, width=0.45\textwidth]{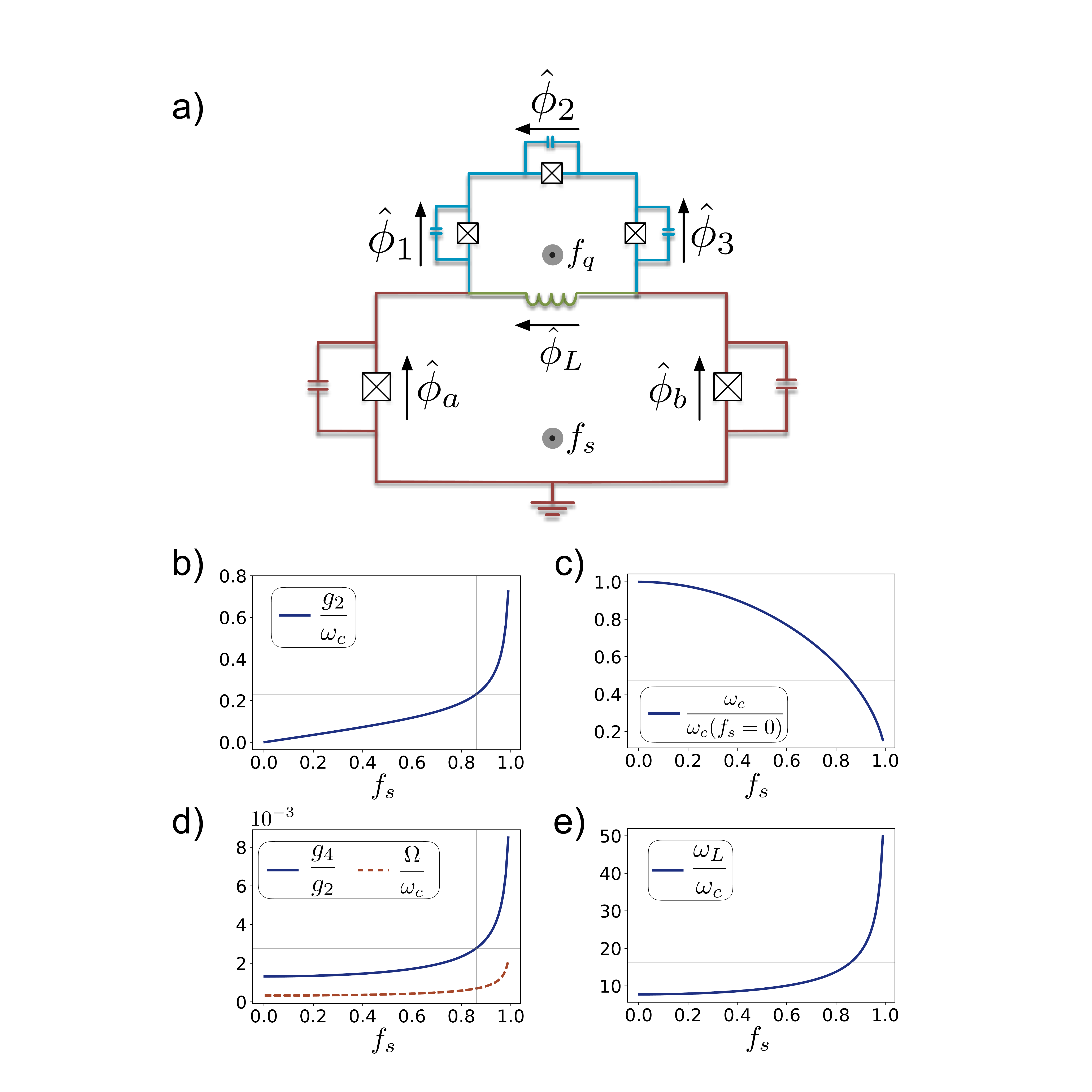}
\caption{\label{chip} (color online) (a) Sketch of the circuit scheme, a flux qubit (cyan) and a  SQUID (red) galvanically coupled through a linear inductive element (green). (b), (c), (d) and (d) Analysis of the system physical parameters as a function of the SQUID-loop flux bias.  Straight grey lines pinpoint the values of parameters for $g_2/\omega_c = 0.23$. For the SQUID and coupling inductance we have set
$E_C = 2\times 10^{-3} E_J$, $ E_L = 30 E_J$. The flux qubit parameters have been chosen in order to meet the two-photon resonance condition $\omega_q = 2\omega_c$. Accordingly, $\widetilde E_J \approx 11.6 E_J$,  $\widetilde E_C=\widetilde E_J/80$ and $\alpha = 0.8$.}
\end{figure}
The proposed circuit is shown in Fig.~\ref{chip}(a). It is composed of a flux qubit and a SQUID, galvanically coupled through a small inductance $L$. The SQUID is operated in the linear regime and, at relevant energy scales, it is well approximated by a quantum harmonic resonator. In the following we show that the nonlinear coupling mediated by the coupling inductance is well approximated by a TPI between the qubit and the resonator mode.
We divide the total circuit Lagrangian in three terms $\mathcal{L} = \mathcal{L}_{\rm SQUID} +  \mathcal{L}_{\rm FQ} + \mathcal{L}_{L}$. By applying the flux-quantization rule, the SQUID Lagrangian can be written as,
\begin{equation}
\label{tobelinearized1}
\mathcal{L}_{\rm SQUID} = C \phidot_+^2  + 2 E_J\Cos{\frac{\varphi_L + f_s}{2}}\Cos{\varphi_+},
\end{equation}
where we defined the symmetric and antisymmetric variables $\varphi_+ = \frac{\varphi_a + \varphi_b}{2}$ and $\varphi_- = \frac{\varphi_a - \varphi_b}{2}$ . The SQUID Josephson junctions have the same Josephson energy $E_J$ and they are shunted by a capacitance $C$.  Here and in the following, $\varphi_i = \phi_i/\phi_0$ is the gauge-invariant phase of the junction $i$. We defined  the superconducting phase difference $\varphi_L$ across the coupling inductance,  the reduced magnetic flux quantum $\phi_0 = \hbar/2e$, and the frustration $f_s = \phi^{\rm ext}_s/\phi_0$ due to a constant magnetic flux threading the SQUID loop.  The term $ \mathcal{L}_{\rm FQ}$ is the standard Lagrangian of a flux qubit~\cite{vanderwal00,Orlando99}, with a modified magnetic bias. The corresponding inductive potential is given by~\cite{SupMat}
\begin{equation}
\label{tobelinearized2}
U =   - 2 \widetilde E_J \Cos{\varphi_p }\Cos{\varphi_m } - \alpha \widetilde E_J  \Cos{2\varphi_m + f},
\end{equation}
where $\widetilde E_J$  is the Josephson energy of the junctions  forming the flux qubit, and $\alpha$ is a dimensionless parameter that renormalizes the parameters of the second junction~\cite{vanderwal00}. We have also defined the flux-qubit symmetric and antisymmetric variables $\phi_p = \frac{\phi_1 + \phi_3}{2}$  and $\phi_m = \frac{\phi_1 - \phi_3}{2}$.
The frustration $f = \varphi_L + f_q $ is the sum of a constant contribution $f_q = \phi^{\rm ext}_q$ due to the flux flowing through the qubit loop and the coupling-inductance phase variable $\varphi_L$. 
Finally, the Lagrangian of the coupling inductance is $\mathcal{L}_{L} =  \frac{C + 2 \alpha \widetilde C}{4} \phidot_L^2 -\frac{\phi^2_L}{2L} $, which corresponds to an LC resonator of frequency $\omega_{\rm L } = \sqrt{\frac{2}{L (C+\alpha\widetilde C )}}$. Notice that in practical implementations the coupling inductance could be replaced by a Josephson junction operated in the linear regime~\cite{Niemczyk:2010}.  The small correction $\widetilde C$ is due to the capacitance of the qubit junctions. Direct inductive coupling between the SQUID and the flux qubit is negligible for typical qubit loops dimensions.

We now take the coupling inductance $L$ to be a small parameter and we adiabatically eliminate $\phi_L$~\cite{SupMat}. 
To simplify the expressions, we define the constants
$K = 2E_J \Cos{\frac{f_s}{2}}$ and  $S= E_J\Sin{\frac{f_s}{2}}$, and the variable $ \Sigma_m =\alpha \widetilde E_J\Sin{2 \varphi_m + f_q}$.¡ 
First, we Taylor-expand Eqs.~\eqref{tobelinearized1} and \eqref{tobelinearized2} up to first order in $\phi_L$.
Then, we assume that $\omega_L$ is much larger than the relevant system frequencies and that  $\dot \phi_L = 0$.  The Euler equation $\partial \mathcal{L} / \partial \phi_L  = 0$ then imposes 
$ \phi_L = -\frac{L}{\phi_0} S\Cos{\varphi_+} - \frac{L}{\phi_0} \Sigma_m$.  
The system Hamiltonian can be derived defining the conjugate variables $p_i = \partial \mathcal{L}_{\rm TOT}/\partial \varphidot_i$ and implementing the corresponding Legendre transformation. Finally, we replace the classical conjugate variables with quantum-mechanical operators $p_i \rightarrow \hat p_i$ and $\varphi_i \rightarrow \hat \varphi_i$, such that $\left[\hat \varphi_i, \hat p_i\right]=i\hbar$.

 Let us discuss the different Hamiltonian terms resulting from this derivation.
In ${\hat H}_{0}$ we gather all noninteracting terms that depend on the SQUID phase variable,
\begin{equation}
\label{SquidHam}
{\hat H}_{0} = \frac{1}{4\phi_0^2 C}\hat p_+^2 -  K \Cos{\hat  \varphi_+} - \frac{S^2}{4E_L}\Cos{\hat  \varphi_+}^2.
\end{equation}
We assume now that the phase of the SQUID junctions is small compared to the magnetic flux quantum $\phi_+/\phi_0\ll 1$. This approximation is valid in the limit of large Josephson energy $E_J\gg E_C= e^2/\left( 2C \right)$. Accordingly, we  expand the cosine functions depending on the phase variable $\hat \varphi_+$, neglecting terms of the order of $\hat \varphi_+^4$. Under this approximation the SQUID Hamiltonian is that of a quantum harmonic oscillator ${\hat H}_{0} =  \hbar \omega_c \hat a^\dagger \hat a  $, where the creation and annihilation operators are defined as $\hat \varphi_+ = \sqrt{\frac{\hbar \omega_c L_{\rm eff}}{2\phi_0^2}}\left( \hat a^\dagger+ \hat a\right)$ and $\hat p_+ = i \sqrt{\frac{\hbar\phi_0^2 }{2\omega_c L_{\rm eff}}}\left( \hat a^\dagger - \hat a\right)$. The frequency of this bosonic mode is given by 
$\omega_c = \sqrt{\frac{1}{  2 C  L_\eff}}$, where $L_\eff = \frac{\phi_0^2}{\left( K + \frac{S^2}{2E_L} \right)}$.

The qubit energy contribution is given by the standard Hamiltonian $ {\hat H}_{\rm FQ}^{\rm standard}$ of a flux qubit~\cite{Orlando99} plus two corrections,
\begin{equation}
\label{QubitHam}
{\hat H}_{\rm FQ} = {\hat H}_{\rm FQ}^{\rm standard} -  \frac{\hat  \Sigma_m^2}{4E_L} -  \frac{S}{4E_L}\hat \Sigma_m.
\end{equation}
The large anharmonicity of the flux qubit allows to perform a two-level approximation, such that ${\hat H}_{\rm FQ}^{\rm standard} = \hbar \omega_{\rm FQ} \hat \sigma_z$. In the two-level subspace we can write~\cite{Liu05} $\hat \Sigma_m \propto \Sin{2 \varphi_m + f_q} \propto \sigma_x$, hence the first correction in Eq.~\eqref{QubitHam} does not couple the qubit ground and excited states. The second correction term $\frac{S}{2E_L}\hat \Sigma_m$ corresponds to a rotation in the qubit basis that can be compensated by a small modification of the qubit flux bias with respect to the sweet-spot $f_q/\phi_0=0.5$.

The interaction Hamiltonian ${\hat H}_I = \frac{S}{4E_L}\hat \Sigma_m \hat \varphi_+^2$ corresponds to a nondipolar interaction between the qubit and the resonator, which is a direct consequence of the nonlinear coupling of Eq.~\eqref{tobelinearized1}. Accordingly, the total system Hamiltonian is well approximated by the  two-photon quantum Rabi model with a full-quadratic coupling,
\begin{equation} 
\label{twoPhRabi}
{\hat H}_{2ph \rm} = \omega_c \hat a^\dagger \hat a + \frac{\omega_q}{2}\hat \sigma_z + g_2 \hat\sigma_x \left( \hat a^\dagger + \hat a  \right)^2.
\end{equation}
The two-photon coupling strength is given by
\begin{equation}
g_2 =  \frac{S}{4E_L}\sqrt{\frac{E_C}{ \left(K + \frac{S^2}{2E_L} \right)}} \bra{0} \hat \Sigma_m \ket{1},
\end{equation}
where $ \ket{0}$ and $ \ket{1}$ are the qubit ground and excited states, respectively. 

Let us now analyze the regimes of parameters accessible with the proposed scheme.
In Fig.~\ref{chip}, we show the dependence of the system parameters on the SQUID flux bias. The effective qubit parameters have been obtained via numerical diagonalization of the Hamiltonian of Eq.~\eqref{QubitHam}. 
As the SQUID flux bias $f_s$ increases, the two-photon coupling strength $g_2$ grows, while the resonator frequency decreases. In Fig.~\ref{chip}(b) we show that the ratio $g_2/\omega_c$ can be brought into the nonperturbative USC regime~\cite{Rossatto17}, making it possible to reach the vicinity of the spectral collapse. Notice that for TPI with full-quadratic coupling~\cite{Felicetti:2018} the collapse takes place for $g_2=0.25\omega_c$. We will take here as a reference the coupling strength $g_2/\omega_c=0.23$.  As we will see in the following, this value is sufficient to observe a clear signature of the spectral collapse. Such value can be achieved with $f_s = 0.86$, where the resonator frequency is approximatively half the value it takes when no flux bias is applied to the SQUID loop $\omega_c = 0.47 
\omega_c^{(f_s=0)}$.

Notice that to obtain Eq.~\ref{twoPhRabi} we have neglected terms of the order $\hat \varphi_+^4$ in the resonator energy and in the coupling Hamiltonian. In Fig.~\ref{chip}(d) we show the ratio between the fourth-order~\cite{SupMat} $g_4$ and the two-photon $g_2$ coupling parameters, and the ratio between the size of the quartic correction $\Omega$ and the resonator frequency $\omega_c$.  Both corrections are three orders of magnitude smaller than the terms considered, until the SQUID flux bias approaches the degeneracy point. On the other hand, the validity of the adiabatic elimination of the coupling inductance  is enhanced for high values of the coupling strength, as shown in Fig.~\ref{chip}(e).

\paragraph{Fluorescence spectrum}
Let us now characterize the response of the system when the cavity or the atom are driven by a monochromatic coherent field and both coupled to a dissipative environment. The total time-dependent Hamiltonian of the system is
\begin{equation}\label{Hamilto}
\hat H(t) = \hat H_{2ph \rm} + F\cos(\omega_dt)(\hat{c}+\hat{c}^{\dagger}),
\end{equation}
where $F$ is the amplitude of the driving field, $\omega_d$ its frequency and $\hat{c}$ is either $\hat{a}$ (cavity driving) or $\hat{\sigma}_-$ (qubit driving).
A Markovian master equation for the density matrix $\rho(t)$ is obtained following a microscopic derivation in the dressed-state basis. The equation has the standard Lindblad form, with jump operators involving transitions between eigenstates of $\hat H_{2ph \rm}$ \cite{Beaudoin:2011, Ridolfo:2012}. The dissipative part reads
\begin{align}\label{MEdiss}
\mathcal{L}\rho =\sum_{p,q =\pm}\sum_{k,j}\Theta(\Delta_{jk}^{pq})\left(\Gamma_{jk}^{pq}+K_{jk}^{pq}\right)\mathcal{D}[|\Psi_j^{p}\rangle\langle \Psi_k^{q}|],
\end{align}
where the quantities $\Gamma_{jk}^{p\bar{p}}$ and $K_{jk}^{p\bar{p}}$ are the rates of transition from a dressed-state $|\Psi_k^{\bar{p}}\rangle$ to $|\Psi_j^{p}\rangle$ due to the atomic and cavity decay, respectively \cite{SupMat}. The variables $p,q \in \{+,-\}$, denote the parity of the number of photons, which is a symmetry of $\hat{H}_{2ph}$. Within a given parity subspace, the eigenstates are labeled in increasing order of the energy, $E^p_j > E^p_i$ for $j>i$. The quantity $\Theta(x)$ is a step function, i.e., $\Theta(x)=0$ for $x\leq0$ and $\Theta(x)=1$ for $x>0$ and $\Delta_{jk}^{pq} = E_k^q - E_j^p$. We have also introduced the notation $\mathcal{D}[\mathcal{O}] = \mathcal{O}\rho\mathcal{O}^{\dagger} - \frac{1}{2}(\rho\mathcal{O}^{\dagger}\mathcal{O} + \mathcal{O}^{\dagger}\mathcal{O}\rho)$.

\begin{figure*}[t]
 	\centering 
	\includegraphics[width = 2\columnwidth]{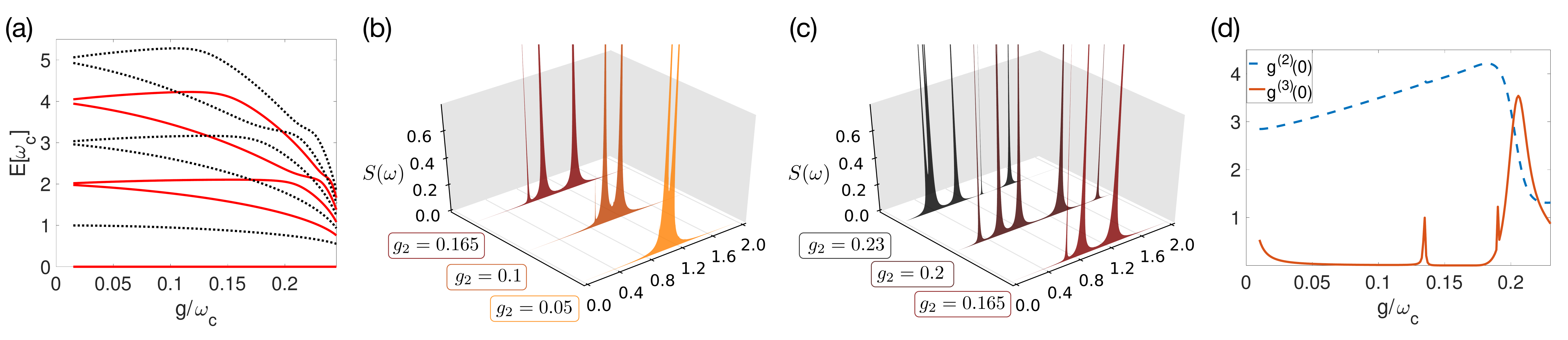}
\caption{(a) Energy spectrum of the two-photon Rabi Hamiltonian (without driving). Red solid lines indicate energy levels with an even number of photons while black dotted lines correspond to an odd number of photons. (b)-(c) Fluorescence spectrum as a function of the coupling strength for $ 0.05\leq g_2/\omega_c \leq 0.165$ inl (b) and $ 0.165\leq g_2/\omega_c \leq 0.23$ in (c). (d) Second and third order autocorrelation function, $g^{(2)}(0)$ and $g^{(3)}(0)$  as a function of $g_2/\omega_c$. The dissipation and driving parameters are $\gamma/\omega_c = \kappa/\omega_q = 10^{-3}$ and $F/\gamma = 1$.}
\label{fig:fluo}
\end{figure*}

We characterize the system through correlation functions of the output field, considering that the resonator is coupled to a one-dimensional waveguide. As shown in Ref.~\cite{Ridolfo:2012}, the output field in the ultrastrong coupling is proportional to an operator $\dot X^+$, defined in the dressed-state basis as
\begin{equation}\label{output}
\dot{X}^+  = \sum_{p=\pm}\sum_{k,j}\Theta(\Delta_{jk}^{p\bar{p}}) \Delta_{jk}^{p\bar{p}}|\Psi_j^{p}\rangle \langle \Psi_j^{p}|i(\hat{a}^\dagger-\hat{a})|\Psi_k^{\bar{p}}\rangle \langle \Psi_k^{\bar{p}}|.
\end{equation}
We first focus on the fluorescence spectrum, extracted from the two-time correlation function $g(t,t+\tau) = \langle \dot{X}^-(t) \dot{X}^+(t+\tau)\rangle$. Given the absence of a rotating frame in which the Hamiltonian is time independent, $g(t,t+\tau)$ depends both on $t$ and $\tau$, but it is periodic in $t$. The Fourier transform (relative to $\tau) $ $S(\omega, t) = \int_{-\infty}^{+\infty} d \tau e^{i\omega \tau}g(t,t+\tau)$ is then also periodic in $t$ and the fluorescence spectrum is given by its zeroth Fourier component \cite{Malz:2016}.
The function $g(t,t+\tau)$ is computed numerically by means of the quantum regression theorem \cite{Carmichael:1998}. In the present case, one efficient way to exploit the quantum regression theorem without performing the numerical integration of the differential equation governing $g(t,t+\tau)$ is to follow a Floquet-Liouville approach \cite{Ho:1986, LeBoite:2017}. Within this framework, all the information about the dynamics of the system is contained in the eigvalues and eigenvector of the Floquet-Liouvillian. For the fluorescence spectrum presented in Fig. (\ref{fig:fluo}), we have checked that both numerical integration and diagonalization of the Floquet-Liouvillian give similar results \cite{SupMat}. 

Before discussing the numerical results, let us recall that, in addition to the parameters of the Hamiltonian, the fluorescence spectrum also depends on the particular driving scheme that we choose (i.e, on the driving frequency $\omega_p$ and the driving amplitude $F/\gamma$). In all what follows, $\omega_p$ is assumed to be resonant with the transition $\ket{\Psi_0^+} \to \ket{\Psi_2^+}$, i.e, from the ground state to the second excited state in the even parity subspace (see the energy spectrum on Fig. \ref{fig:fluo}). Note that coupling two states with the same parity is possible only when driving the qubit ($\hat{c} = \hat{\sigma}_-$ in Eq. (\ref{Hamilto})).  As we will see shortly, this driving scheme is well suited to capture two main features of the USC regime : (i) the breaking of the selection rules and the change in symmetry due to counterrotating terms (ii) the onset of the spectral collapse for $g_2/\omega_c \to 0.25$. Another important point is that the fluorescence photons that we consider result from the emission of resonator photons into the  output waveguide. Therefore, as shown in Eq. (\ref{output}), only the operators $a$ and $a^{\dagger}$ enter in the definition of the output field, which means that only transitions changing the parity contribute to the fluorescence spectrum. In particular $\ket{\Psi_2^+} \to \ket{\Psi_0^+}$ and $\ket{\Psi_2^+} \to \ket{\Psi_1^+}$ are excluded.

Fluorescence spectrums for different values of the coupling strength, ranging from $g_2/\omega_c = 0.05$ to $g_2/\omega_c = 0.23$ are presented in Fig. \ref{fig:fluo} (b) and (c). For the sake of clarity they are separated into two parts. In Fig. \ref{fig:fluo} (b), we observe the breaking of the selection rules and the approximate symmetry due to the rotating wave approximation (RWA) as one enters the USC regime. For $g_2/\omega_c = 0.005$ the spectrum has only two peaks corresponding to the transitions $\ket{\Psi_2^+} \to \ket{\Psi_0^-}$ and $\ket{\Psi_0^-} \to \ket{\Psi_0^+}$, which is what we expect in the regime where RWA is valid. Indeed, the RWA implies the conservation of the weighted excitation number $2a^{\dagger}a + \sigma_z$ \cite{Felicetti:2018} and vanishing matrix elements $\bra{\Psi_2^+}a\ket{\Psi_1^+} = \bra{\Psi_2^+}\sigma_-\ket{\Psi_1^+} = 0 $. This is no longer the case when one increases the coupling strength and enters the USC regime. A third resonance at the frequency of the transition $\ket{\Psi_1^+} \to \ket{\Psi_0^-}$ appears for $g_2/\omega_c = 0.1$ and $g_2/\omega = 0.165$, which means that the transition $\ket{\Psi_2^+} \to \ket{\Psi_1^+}$ is no longer forbidden by selection rules. In other words, the approximate RWA symmetry is not valid anymore and counter-rotating terms start to play an important role in the dynamics.   

The structure of the fluorescence spectrum changes drastically when the coupling strength is increased further. As seen in Fig.~\ref{fig:fluo} (c),  multiple additional peaks emerge for $g_2/\omega_c = 0.2$. This feature is related to a level crossing occurring in the energy spectrum of the two-photon QRM for $g_2 = g_{\rm cross} \approx 0.17$ (See Fig.~\ref{fig:fluo} (a)). For $g_2>g_{\rm cross}$, the energy of the driven state $E_2^+$ becomes higher than $E_1^-$, which implies that many different paths leading to the emission of output photons are now allowed when going from $\ket{\Psi_2^+}$ to the ground state. More importantly,  for $g_2/\omega_c = 0.23$, the resonances appearing in the spectrum are globally red-shifted, i.e. the same number of resonances is spread over a smaller interval. The highest frequency for example shifts from $1.8\omega_c$ ($g_2/\omega_c = 0.2$) to $1.3\omega_c$ ($ g_2/\omega_c = 0.23$). This gets more and more pronounced as the coupling strength tends to $0.25\omega_c$ and is a clear signature of the onset of the spectral collapse predicted for the two-photon QRM.

\paragraph{Two-photon blockade}
The fluorescence spectrum for $g_2<g_{\rm cross}$ is also a signature of a two-photon analogue of the celebrated photon blockade effect \cite{Hamsen:2017}. Namely, in the fluorescence spectrum for $g_2/\omega_c = 0.1$,  the three resonances are the signature of two decay channels, $\ket{\Psi_2^+} \to \ket{\Psi_0^-} \to \ket{\Psi_0^+}$ and $\ket{\Psi_2^+} \to \ket{\Psi_1^+} \to \ket{\Psi_0^-} \to \ket{\Psi_0^+}$. As $\ket{\Psi_2^+} \to \ket{\Psi_1^+}$ does not give rise to the emission of an output photon, these channels result in emissions of single  photons or photon pairs. In addition, the anharmonicity of the spectrum ensures that no higher-oder transitions are resonant with the driving frequency. More insight into this phenomenon can be gained by computing the second- and third-order autocorrelation functions of the output field defined as $g^{(2)}(0) = {\langle \dot X^-\dot X^-\dot X^+ \dot X^+\rangle}/{\langle \dot X^-\dot X^+\rangle^2}$ and $g^{(3)}(0) = {\langle (\dot X^-)^3(\dot X^+)^3\rangle}/{\langle \dot X^-\dot X^+\rangle^3}$. The two-photon blockade is characterized by ``two-photon" antibunching, i.e., the conjunction of $g^{(2)}(0) > 1$ and $g^{(3)}(0) < 1$.  As shown in Fig. \ref{fig:fluo} (d), we always have $g^{(2)}(0) > 1$ for the driving scheme we consider. On the other hand, $g^{(3)}(0)$ shows very rich higher-order photon correlations due to the TPI. First, focusing on the global behaviour, we observe that the output field shows strong two-photon antibunching in the USC regime until $g_2  = g_{\rm cross}$. The increase in $g^{(3)}(0)$ for $g_2>g_{\rm cross}$ is a direct consequence of the level crossing in the energy spectrum and the multiple decay channels available after this point. Second, it shows sharp peaks at  $g_2/\omega \approx 0.13$ and $g_2/\omega \approx 0.19$. This suppression of the two-photon antibunching occurs because the energy spectrum becomes less anharmonic at these points, allowing transitions to higher energy levels.
 
\paragraph{Conclusions} We have proposed a superconducting quantum circuit to explore experimentally the physics of TPI in the nonperturbative USC regime. The fluorescence spectrums and output fields correlation functions reveal fundamental differences with respect to standard dipolar interactions. We have identified a driving protocol that allows one to characterize the transition form strong to ultrastrong coupling,  to reveal rich higher-order photon correlations and to observe a clear signature of the spectral collapse.

\begin{acknowledgments}
We acknowledge insightful discussions with Pol Forn-D\'iaz, Juan Jos\'e Garcia-Ripoll and Daniel Braak.
S.F. acknowledges support from the French Agence Nationale de la Recherche (SemiQuantRoom, Project No. ANR14-CE26-0029) and from the PRESTIGE program, under the Marie Curie Actions-COFUND of the FP7. M.J.H. was supported by the ERC Synergy grant BioQ.
\end{acknowledgments}

\pagebreak
\widetext
\begin{center}
\textbf{ \large Supplemental material for Ultrastrong coupling regime of non-dipolar light-matter interactions}
\end{center}
\setcounter{equation}{0}
\setcounter{figure}{0}
\setcounter{table}{0}

\renewcommand{\theequation}{S\arabic{equation}}
\renewcommand\thefigure{S\arabic{figure}}    
\renewcommand\thesection{S\arabic{section}}
\renewcommand\thesubsection{S\arabic{section}.\arabic{subsection}}
\renewcommand\theparagraph{\bf S\arabic{section}.\arabic{paragraph}}

\def\ln{{\rm ln}}
\def\s{{\rm\bf \sigma}}
\def\x{{\rm\bf x}}
\def\y{{\rm\bf y}}
\def\p{{\rm\bf p}}
\def\q{{\rm\bf q}}
\def\k{{\rm\bf k}}
\def\U{{\rm U}}
\def\l{{\it  l}}
\def\d{{\rm d}}
\def\la{\langle}
\def\ra{\rangle}
\def\om{\omega}
\def\Om{\Omega}
\def\vep{\varepsilon}
\def\wh{\widehat}
\def\tr{{\rm Tr}}
\def\da{\dagger}
\def\beq{\begin{equation}}
\def\eeq{\end{equation}}

\section{A.\: Circuit model}
\label{SM_sec1}
In this section we provide a detailed derivation of the circuit model presented in the main text.
\subsection{A.1.\: Lagrangian}
The system Lagrangian is given by:
\begin{eqnarray}
\mathcal{L}_{\rm SQUID} &=& \frac{C}{2}\dot \phi_a^2 +  \frac{C}{2}\dot \phi_b^2 +  E_J\left[\Cos{\frac{\phi_a}{\phi_0}} + \Cos{\frac{\phi_a}{\phi_0}}  \right] \\ 
\mathcal{L}_{\rm FQ} &=& \frac{\widetilde C}{2} \left[\phidot_1^2 + \phidot_3^2 \right] +  \frac{\alpha \widetilde C}{2}\phidot_2^2 
+ \widetilde{E}_J \left[\Cos{\frac{\phi_1}{\phi_0}} + \Cos{\frac{\phi_3}{\phi_0}} + \alpha \Cos{\frac{\phi_2}{\phi_0}}  \right] \\
\label{SM_lagL}
\mathcal{L}_{\rm L} &=& -\frac{\phi^2_L}{2L}.
\end{eqnarray}
We defined the reduced magnetic flux quantum as $\phi_0 = \Phi_0/2\pi = \hbar/2e$. 
We define symmetric and anti-symmetric SQUID variables as
$\phi_+ = \frac{\phi_a + \phi_b}{2}$ and $\phi_- = \frac{\phi_a - \phi_b}{2}$. 
Applying the flux-quantization rules for the SQUID loop $\phi_a - \phi_b = \phi_L + \phi^{\rm ext}_s $, we can express the anti-symmetric variable in terms of the coupling inductance phase and the external flux flowing through the SQUID loop  $\phi_- = \frac{\phi_L}{2} + \frac{\phi^{\rm ext}_s}{2}$. Assuming a constant external flux $\dot \phi^{\rm ext}_s=0$, simple trigonometric relations allow to rewrite the SQUID Lagrangian as,
\begin{equation}
\label{SM_tobelinearized1}
\mathcal{L}_{\rm SQUID} = C \phidot_+^2 + \frac{C}{4} \phidot_L^2 + 2 E_J\Cos{\frac{\phi_L + \phi^{\rm ext}_s}{2\phi_0}}\Cos{\frac{\phi_+}{\phi_0}}.
\end{equation}

We define symmetric and antisymmetric variables for the flux qubit $\phi_p = \frac{\phi_1 + \phi_3}{2}$  and $\phi_m = \frac{\phi_1 - \phi_3}{2}$. The  flux-quantization rule for the qubit loop $\phi_1 - \phi_2 - \phi_3 = -  \phi_L - \phi^{\rm ext}_q $ allows to eliminate the phase variable of the smaller junction $\phi_2 = 2\phi_m + \phi_L + \phi^{\rm ext}_q$. Notice that $ \phi^{\rm ext}_s$ is defined in the opposite direction with respect to $ \phi^{\rm ext}_q$. We will assume that the classical flux biasing the flux qubit is also constant $\dot \phi^{\rm ext}_q = 0$, so that
\begin{equation}
\label{SM_tobelinearized2}
\mathcal{L}_{\rm FQ} =  \widetilde{C} \phidot_p^2 + \widetilde{C} \phidot_m^2 + \frac{\alpha\widetilde C}{2}\left( 2\phidot_m + \phidot_L \right)^2 + 
\widetilde{E}_J \left[ 2\Cos{\frac{\phi_p}{\phi_0} }\Cos{\frac{\phi_m}{\phi_0} } + \alpha \Cos{\frac{2\phi_m + \phi_L +  \phi^{\rm ext}_q}{\phi_0}}  \right].
\end{equation}

In the following we  take a perturbative approach considering the coupling inductance $L$ as a small parameter. Accordingly, we perform two approximations: we linearize the Lagrangian with respect to the coupling-inductance phase variable $\phi_L$, and then we adiabatically eliminate the corresponding degree of freedom. To simplify the expressions, we define gauge-invariant phase variables $\varphi_i= \phi_i/\phi_0$ and the frustrations $f_s = \phi^{\rm ext}_s/\phi_0$ and $f_q = \phi^{\rm ext}_q/\phi_0$.
\paragraph{Linearization --}
 We assume that the flux variable $\phi_L$ is small with respect to the magnetic flux quantum, and so we linearize Eq.\eqref{SM_tobelinearized1} with respect to $\phi_L$,
\begin{equation}
\label{SM_lagSQUID}
\mathcal{L}_{\rm SQUID} = C \phidot_+^2 + \frac{C}{4} \phidot_L^2 +
 2 E_J \left[\Cos{\frac{f_s}{2}} - \Sin{f_s/2}\frac{\varphi_L}{2} \right] \Cos{\varphi_+}.
\end{equation}
We also linearize Eq.\eqref{SM_tobelinearized2} with respect to $\phi_L$,
\begin{equation}
\label{SM_lagFQ}
\mathcal{L}_{\rm FQ} =  \mathcal{L}_{\rm qubit} + \frac{\alpha\widetilde C}{2}\left(\phidot_L^2 + 4 \phidot_L\phidot_m \right) 
- \frac{\alpha \widetilde{E}_J}{\phi_0} \Sin{2\varphi_m+f_q}\phi_L,
\end{equation}
where with $ \mathcal{L}_{\rm qubit}$ we denote the standard Lagrangian of a flux qubit~~\cite{vanderwal00,Orlando99},
\begin{equation}
 \mathcal{L}_{\rm qubit} = \widetilde{C} \phidot_p^2 + \left(1+2\alpha \right)\widetilde{C}  \phidot_m^2 + 
E_J \left[ 2\Cos{\varphi_p }\Cos{\varphi_m } + \alpha \Cos{2\varphi_m + f_q}  \right].
\end{equation}

\paragraph{Adiabatic elimination --}
Let us now focus on the branch variables relative to the coupling inductance, the free Lagrangian term in $(\phi_L, \phidot_L)$ is given by
\begin{equation}
\mathcal{L}_{\rm coupler} = \left(\frac{C+2\alpha\widetilde C}{2}\right) \frac{\phidot^2_L}{2} - \frac{\phi_L^2}{2L},
\end{equation}
which corresponds to a harmonic oscillator of frequency $\omega_{\rm L } = \sqrt{\frac{2}{L (C+\alpha\widetilde C )}}$. $L$ being a small parameter in our development, we assume that $\omega_L$ is much larger than all other system frequencies, and we adiabatically eliminate  the corresponding degree of freedom imposing $\phidot_L = 0$. In order to simplify the expressions, in the following we will make use of the following parameters
\begin{equation}
\label{SM_pardef}
K = 2E_J \Cos{\frac{f_s}{2}}\ ; \quad S= E_J\Sin{\frac{f_s}{2}}\ ; \quad \Sigma_m =\alpha \widetilde E_J\Sin{2\varphi_m + f_q}.
\end{equation}
From Euler's equation we obtain the dependence of $\phi_L$ on the remaining dynamic variables,
\begin{equation}
\label{SM_phiL}
\frac{\partial \mathcal{L}_{\rm TOT}}{\partial \phi_L} = 0 \longrightarrow \phi_L = -\frac{L}{\phi_0} S\Cos{\varphi_+} - \frac{L}{\phi_0} \Sigma_m.
\end{equation}

The total Lagrangian is obtained adding the equations  \eqref{SM_lagSQUID}, \eqref{SM_lagFQ} and  \eqref{SM_lagL}, 
$\mathcal{L}_{\rm TOT} =\mathcal{L}_{\rm SQUID} + \mathcal{L}_{\rm FQ} + \frac{\phi_L^2}{2L}$, and it can be written as
\begin{equation}
\label{SM_LagphiL}
\mathcal{L}_{\rm TOT} = C\phidot^2_+ + \left[ K- S \varphi_L \right] \Cos{\varphi_+} +  \mathcal{L}_{\rm qubit} -  \Sigma_m \varphi_L - \frac{\phi_0^2}{2L}\varphi_L^2.
\end{equation}
By replacing $\varphi_L$ with Eq.~\eqref{SM_phiL}, we obtain finally

\begin{equation}
\mathcal{L}_{\rm TOT} =  \textcolor{BrickRed}{C\phidot^2_+ + K \Cos{\varphi_+} + \frac{S^2}{4E_L}\Cos{\varphi_+}^2 } + \textcolor{MidnightBlue}{\mathcal{L}_{\rm qubit} + \frac{\Sigma_m^2}{4E_L}} 
+ \frac{S}{2E_L}\Sigma_m \Cos{\varphi_+},
\end{equation}
where we defined the inductance energy $E_L = \phi_0^2/2L$, and we highlighted in red and blue the free energy terms of the SQUID and the flux qubit, respectively.

\subsection{A.2.\: Hamiltonian}
The system Hamiltonian can be finally derived defining standard conjugate variables $p_i = \partial \mathcal{L}_{\rm TOT}/\partial \varphidot_i$ and implementing the Legendre transformation. We replace now the classical variables with quantum operators and we start using the hat formalism to avoid confusion. The Hamiltonian can be written as
\begin{equation}
\label{SM_fullHam}
\mathcal{\hat H} = \mathcal{\hat H}_{\rm SQUID} + \mathcal{\hat H}_{\rm FQ} + \mathcal{\hat H}_I,
\end{equation}
let us discuss the three terms independently.
Using the parameters defined in Eq.~\eqref{SM_pardef}, the SQUID Hamiltonian is given by
\begin{equation}
\label{SM_SquidHam}
\mathcal{\hat H}_{\rm SQUID} = \frac{\hat p_+^2}{4C\phi_0^2} -  K \Cos{\hat  \varphi_+} - \frac{S^2}{4E_L}\Cos{\hat  \varphi_+}^2.
\end{equation}

The Hamiltonian $\mathcal{\hat H}_{\rm FQ}$ is given by the standard flux-qubit Hamiltonian, plus a correction proportional to the small parameter $L$,
\begin{eqnarray}
\label{SM_fluxqubit_first}
\mathcal{\hat H}_{\rm FQ} &=&  \frac{\hat p_p^2}{4 \widetilde C \phi_0^2 } +  \frac{\hat p_m^2}{4\widetilde C \phi_0^2  (1+2\alpha)}  + \\ 
&\ & - E_J \left[ 2\Cos{\hat \varphi_p }\Cos{ \varphi_m } + \alpha \Cos{2\hat \varphi_m + fq}  \right] -  \frac{\hat  \Sigma_m^2}{4E_L} \nonumber
\end{eqnarray}
The last term in Eq.~\eqref{SM_fullHam} corresponds to the nondipolar coupling Hamiltonian.
\begin{equation}
\label{SM_nondipcoup_1}
\mathcal{\hat H}_I = -\frac{S}{2E_L}\hat \Sigma_m \Cos{\hat \varphi_+}.
\end{equation}
We show in the following that, in a broad regime of parameters, such nondipolar coupling can be reduced to a two-photon interaction plus an additional correction to the flux-qubit Hamiltonian.

\subsection{A.3.\: Effective model}
Let us now assume that the phase of the SQUID junctions is small compared to the magnetic flux quantum $\varphi_+ = \phi_+/\phi_0\ll 1$. This approximation is valid in the limit of large Josephson energy. 
Expanding the cosines in Eq.~\eqref{SM_SquidHam} we obtain,
\begin{equation}
\label{SM_SQUIDexp}
\mathcal{\hat H}_{\rm SQUID} = \frac{\hat p_+^2}{4\phi_0^2C} +  \left(K + \frac{S^2}{2E_L} \right)  \frac{\hat  \varphi_+^2}{2}  -  \left(K + \frac{2S^2}{E_L} \right)  \frac{\hat  \varphi_+^4}{24},
\end{equation}
We kept orders up to $\hat \varphi_+^4$ and we discarded constant terms. Similarly, from Eq.~\eqref{SM_nondipcoup_1}
\begin{equation}
\label{SM_nondipcoup_2}
\mathcal{\hat H}_I = - \frac{S}{2E_L}\hat \Sigma_m + \frac{S}{2E_L}\hat \Sigma_m \left( \frac{\hat  \varphi_+^2}{2} -  \frac{\hat  \varphi_+^4}{24} \right),
\end{equation}
where the first term is a free energy term of the qubit, while the second term is the origin of the nondipolar coupling.

We now define the standard ladder operators of the quantum Harmonic oscillator corresponding to the quadratic part of the SQUID Hamiltonian in Eq.~\eqref{SM_SQUIDexp},
\begin{equation}
\label{SM_def_a_adag}
\hat \varphi_+ = \sqrt{\frac{\hbar \omega_c L_{\rm eff}}{2\phi_0^2}}\left( \hat a^\dagger+ \hat a\right),\quad 
\hat p_+ = i \sqrt{\frac{\hbar \phi_0^2}{2\omega_c L_{\rm eff}}}\left( \hat a^\dagger - \hat a\right),
\end{equation}
where we defined
\begin{equation}
L_\eff = \frac{\phi_0^2}{\left( K + \frac{S^2}{2E_L} \right)},  \ \quad
\omega_c = \sqrt{\frac{1}{  2 C  L_\eff}} = \frac{1}{\hbar}\sqrt{4E_C\left(K + \frac{S^2}{2E_L} \right)},
\end{equation}
and where we introduced  the charging energy $E_C = e^2/2C$.
Equation Eq.~\eqref{SM_SQUIDexp} can be then rewritten as
\begin{equation}
\label{SM_rabi1}
\mathcal{\hat H}_{\rm SQUID} = \hbar \omega \hat a^\dagger \hat a 
-  \Omega \left(\hat a^\dagger + \hat a \right)^4, \quad \text{with}\ \Omega = \frac{E_C\left(K + \frac{2S^2}{E_L} \right)}{24\left(K + \frac{S^2}{2E_L} \right)},
\end{equation}
where $\Omega$ is the size of the first nonlinear correction to the harmonic approximation of the SQUID Hamiltonian.

The total qubit Hamiltonian is given by  Eq.~\eqref{SM_fluxqubit_first} plus  the free term in Eq.~\eqref{SM_nondipcoup_2}
\begin{equation}
\label{SM_rabi2}
\mathcal{\hat H}_{\rm FQ} = \mathcal{\hat H}_{\rm FQ}^{\rm standard} -  \frac{\hat  \Sigma_m^2}{4E_L} -  \frac{S}{2E_L}\hat \Sigma_m ,
\end{equation}
which corresponds to the standard Hamiltonian of a flux qubit plus two corrections. In the qubit subspace we can write 
$\hat \Sigma_m = \alpha \widetilde E_J \bra{0} \Sin{2\hat \varphi_m + f_q} \ket{1} \hat \sigma_x = \alpha \widetilde E_J T(f_q) \hat \sigma_x $, therefore the first correction corresponds to a constant energy offset while the second one can be compensated by a small adjustment  of the frustration $f_q$.

Finally, the second term in Eq.~\eqref{SM_nondipcoup_2} corresponds to the nondipolar interaction Hamiltonian between the qubit and the resonator mode,
\begin{equation}
\label{SM_rabi3}
\mathcal{\hat H}_I = g_2 \sigma_x \left(\hat a^\dagger + \hat a \right)^2 + g_4 \sigma_x \left(\hat a^\dagger + \hat a \right)^4
\end{equation}
where we defined the two- and four-photon coupling strengths $g_2$ and $g_4$, respectively.
\begin{equation}
g_2 =  \frac{S}{4E_L}\sqrt{\frac{E_C}{ \left(K + \frac{S^2}{2E_L} \right)}} \alpha \widetilde E_J T(f_q), \quad g_4 = \frac{S}{48 E_L}\frac{E_C}{  \left(K + \frac{S^2}{2E_L} \right)} \alpha \widetilde E_J T(f_q).
\end{equation}

To conclude, the total system Hamiltonian up to fourth order in $\hat \phi_+$ is given by the sum of equations \eqref{SM_rabi1}, \eqref{SM_rabi2} and \eqref{SM_rabi3}. In the main text we report the results of numerical simulations of the system Hamiltonian, showing that in a large region of physical parameters the quartic corrections $\Omega$ and $g_4$ are negligibly small. Accordingly, the system Hamiltonian is faithfully approximated by the two-photon quantum Rabi model. Notice that the higher-order contributions will become relevant to renormalize the spectrum of the physical model once the spectral collapse takes place.


\section{B.\: Master equation in the dressed-state basis}

In this section we provide additional information on the derivation of the master equation. Following the work of Ridolfo \textit{et al} \cite{Ridolfo:2012}, we assume that dissipation occurs via the coupling of the system to one-dimensional waveguides, described by the following Hamiltonian 
\begin{equation}\label{SBham}
H_{SB} \propto \int d\omega \sqrt{\omega}(\hat{c}-\hat{c}^{\dagger})(\hat{b}_\omega- \hat{b}^{\dagger}_\omega),
\end{equation}
where  $\hat{c} \in \{\hat{a}, \hat{\sigma}_-\}$ and $\hat{b}_\omega$ are annihilation operators relative to the waveguide modes. The derivation of the master equation follows the microscopic approach in the weak-coupling limit. It is expressed in the eigenbasis $\{ \ket{\Psi_{j}^p}\}$ of the system Hamiltonian $\hat{H}_{2phot}$.  In labelling the eigenstates  we have introduced the variable $p \in \{+,-\}$ that denotes the parity of the number of photons, which is a symmetry of the Hamiltonian. Within a given parity subspace, the eigenstates are labeled in increasing order of the energy, $E^p_j > E^p_i$ for $j>i$.

Under these assumptions the dissipative part of the master equation takes a standard Lindblad form, which reads
\begin{align}\label{MEdiss}
\mathcal{L}\rho =\sum_{p,q =\pm}\sum_{k,j}\Theta(\Delta_{jk}^{pq})\left(\Gamma_{jk}^{pq}+K_{jk}^{pq}\right)\mathcal{D}[|\Psi_j^{p}\rangle\langle \Psi_k^{q}|],
\end{align}
The quantity $\Theta(x)$ is a step function, i.e., $\Theta(x)=0$ for $x\leq0$ and $\Theta(x)=1$ for $x>0$ and $\Delta_{jk}^{pq} = E_k^q - E_j^p$. We have also introduced the notation 
\begin{equation}
\mathcal{D}[\mathcal{O}] = \mathcal{O}\rho\mathcal{O}^{\dagger} - \frac{1}{2}(\rho\mathcal{O}^{\dagger}\mathcal{O} + \mathcal{O}^{\dagger}\mathcal{O}\rho).
\end{equation}
The quantities $\Gamma_{jk}^{pq}$ and $K_{jk}^{pq}$ denote the rates of transition from a dressed-state $|\Psi_k^{q}\rangle$ to $|\Psi_j^{p}\rangle$ due to the atomic and cavity decay, respectively.  They read
\begin{align}\label{trans_rates}
\Gamma_{jk}^{pq} &= \gamma\frac{\Delta_{jk}^{p q}}{\omega_c}|\langle \Psi_j^{p}|(\hat{a} - \hat{a}^{\dagger})|\Psi_k^{q}\rangle|^2, \\
 K_{jk}^{pq} &= \kappa\frac{\Delta_{jk}^{pq}}{\omega_q}|\langle \Psi_j^{p}|(\hat{\sigma}_- - \hat{\sigma}_+)|\Psi_k^{q}\rangle|^2,
\end{align}
where $\Delta_{jk}^{pq} = E_k^{q} - E_j^{p}$ is the transition frequency and $\gamma$, $\kappa$ are respectively the cavity and the atom decay rates.
 
Note that the system Hamiltonian on which the derivation is based does not include the external driving. For the values considered in the main text, the driving is sufficiently weak not to change the form of the jump operators (there is no ``dressing of the dressed-states" by the external field).

\section{C.\: Fluorescence spectrum within Floquet-Liouville theory}
We derive in this section a semi-analytical expression for the fluorescence spectrum within the framework of Floquet-Liouville theory.
\subsection{C.1.\: Floquet-Liouville propagator}
Due to the external driving field, the total Hamlitonian is $T$-periodic, with $T = 2\pi/\omega_d$. The Floquet-Liouville approach allows to get rid of the explicit time-dependence of the master equation by reformulating the dynamics in the space $\mathrm{Op}(\mathcal{H})\otimes \mathcal{T}$, of time-periodic operators, where $\mathcal{H}$ is the underlying Hilbert space of physical states and $\mathcal{T}$ is the space of periodic functions. Using the following convention for Fourier series of periodic functions, $f(t) = \sum_{n = -\infty}^{\infty} f^{(n)}e^{-i n\omega_d t}$, one can define a scalar product on $\mathrm{Op}(\mathcal{H})\otimes \mathcal{T}$ as
\begin{equation}\label{scalarProd}
\langle \langle A|B\rangle\rangle = \sum_n \mathrm{Tr}[A^{(n) \dagger}B^{(n)}],
\end{equation}
which derives from the usual scalar product on $\mathcal{T}$, $(f|g) = \frac{1}{T}\int_{0}^T f^*(t)g(t)\mathrm{d}t$ and the scalar product on $\mathrm{Op}(\mathcal{H})$, $\langle A|B\rangle = \mathrm{Tr[A^{\dagger}B}]$. Within this framework the dynamics is encoded in the eigenvalues and eigenvectors of the Floquet-Liouville superoperator $\mathscr{L}$, which is time independent. For a complete derivation see Refs. \cite{Ho:1986} and \cite{LeBoite:2017}. The central eigenvalue problem is then

\begin{equation}\label{eigProbFlo}
{\mathscr{L}}|R_{\alpha,k}\rangle \rangle = \Omega_{\alpha,k}|R_{\alpha,k}\rangle \rangle,
\end{equation}
where $0\leq \alpha \leq \mathrm{dim}(\mathcal{H})^2-1$, and $k \in \mathbb{Z}$. Note that the object $|R_{\alpha,k}\rangle\rangle$ defines a periodic matrix, i.e.  for each time $t$, $R_{\alpha,k}(t) \in \mathrm{Op}(\mathcal{H})$. As $\mathscr{L}$ is not Hermitian, it is necessary to distinguish the right eigenvectors defined above from the left eigenvectors obeying
\begin{align}
\mathscr{L}^{\dagger}|L_{\alpha , k} \rangle \rangle &= \Omega^*_{\alpha , k}|L_{\alpha , k} \rangle \rangle.
\end{align}
Using these notations, one can write a propagator for the master equation
\begin{equation}
\rho(t+\tau) = U(t+\tau,t)[\rho(t)] = \sum_{\alpha,k} e^{-i\Omega_{\alpha,k}\tau}\langle\langle L_{\alpha,k}|\rho\rangle\rangle R_{\alpha,k}(t+\tau)
\end{equation}

%
\subsection{A.2.\: Quantum regression theorem and fluorescence spectrum}
In the ultrastrong coupling regime, there is no rotating frame in which the Hamiltonian is time-independent. As a result, the correlation function
\begin{equation}\label{2timeCorr}
g(t,t+\tau) = \langle \dot{X}^-(t) \dot{X}^+(t+\tau)\rangle
\end{equation}
depends both on $t$ and $\tau$. In the present context, due to the periodic driving, it is periodic in $t$. We can therefore define a periodic spectrum $S(\omega, t) = \int_{-\infty}^{+\infty} e^{i\omega \tau}g(t,t+\tau)$.
Following the derivation of the standard Wiener-Khinchin theorem, we find that the relevant quantity is the zeroth Fourier component of $S(\omega,t)$
\begin{equation}\label{fluoDef}
S(\omega) = \lim_{t_0\to \infty }\left[\frac{1}{T}\int_{t_0}^{t_0+T} \int_{-\infty}^{+\infty}e^{i\omega \tau}g(t,t+\tau)dt d\tau\right],
\end{equation}
The function in Eq. (\ref{2timeCorr}) is computed by applying the quantum regression theorem. For any operators $\hat{a}$, $\hat{b}$ and time $\tau > 0$  we have
\begin{align}
\langle a(t) b(t+\tau)\rangle  &=  \mathrm{Tr}\{bU(t+\tau,t)[\rho(t)a]\}.
\end{align}
Injecting the expression of the propagator into this last expression we find
\begin{align}
\langle \dot{X}^-(t) \dot{X}^+(t+\tau)\rangle &= \sum_{\alpha,k} e^{-i\Omega_{\alpha,k}\tau}\langle\langle L_{\alpha,k}|\rho_{\infty}\dot X^-\rangle\rangle \mathrm{Tr}[\dot X^+ R_{\alpha,k}(t+\tau)]\\
& = \sum_{\alpha,k,m} e^{-i(\Omega_{\alpha,k}+m\omega_d)\tau -im\omega_d t}\langle\langle L_{\alpha,k}|\rho_{\infty}\dot X^-\rangle\rangle \mathrm{Tr}[\dot X^+ R^{(m)}_{\alpha,k}]. \label{corrFun}
\end{align}
Explicit calculation for $\tau < 0$ can be avoided by inverting the order of  the integrals in Eq. (\ref{fluoDef}). The spectrum then reads  $S(\omega) = 2 \mathrm{Re}[\int_0^\infty d\tau e^{i\omega\tau} g_+(\tau)]$, with $g_+(\tau) = \lim_{t_0 \to \infty} \frac{1}{T}\int_{t_0}^{t_0+T}g(t,t+\tau) dt$.
Since averaging on $t$ selects the component $m = 0$ in Eq. (\ref{corrFun}), we find
\begin{align}
g_{+}(\tau) &= \sum_{\alpha,k} e^{-i\Omega_{\alpha,k}\tau} \langle\langle \dot X^-,0| R_{\alpha,k}\rangle\rangle\langle\langle L_{\alpha,k}|\rho_{\infty}\dot X^-\rangle\rangle
\end{align}
The final semi-analytical expression for the fluorescence spectrum is then
\begin{equation}
S(\omega) = - 2 \mathrm{Re}\left[ \frac{\langle\langle \dot X^-,0| R_{\alpha,k}\rangle\rangle\langle\langle L_{\alpha,k}|\rho_{\infty}\dot X^-\rangle\rangle}{i(\omega -\Omega_{\alpha,k})}\right].
\end{equation}
Due to the dissipative nature of the system, the eigenvalues $\Omega_{\delta \eta,k}$ are complex and satisfy $\mathrm{Im}[\Omega_{\alpha,k}]\leq 0$. If $\mathrm{Im}[\Omega_{\alpha,k}] = 0$, the denominator has to be understood as
\begin{equation}
\frac{-1}{i(\omega-\Omega_{\alpha,k}+i0^+)}  = \pi\delta(\omega-\Omega_{\alpha,k})+ i\mathcal{P}\frac{1}{\omega-\Omega_{\alpha,k}},
\end{equation}
where $\mathcal{P}$ stands for Cauchy's principal value.

\end{document}